\documentclass[11pt]{article}
\usepackage{graphicx}
\usepackage{color}

\providecommand{\e}[1]{\ensuremath{\times 10^{#1}}}

\begin{document}

\title{\textbf{Selective tuning of high-$Q$ silicon photonic crystal nanocavities via laser-assisted local oxidation}}

\author{Charlton J. Chen,$^{1,*}$ Jiangjun Zheng,$^{1,*}$ Tingyi Gu,$^{1}$ James F. McMillan,$^{1}$ \\
Mingbin Yu,$^{2}$ Guo-Qiang Lo,$^{2}$ Dim-Lee Kwong,$^{2}$ and Chee Wei Wong$^{1,**}$ \\ \\
\emph{$^{1}$ Optical Nanostructures Laboratory, Columbia University, New York 10027} \\
\emph{$^{2}$ The Institute of Microelectronics, 11 Science Park Road, Singapore Science}\\
\emph{Park II, Singapore 117685, Singapore}\\ \\
$^{*}$ These authors contributed equally to this work. \\
{\color{blue}$^{**}$ \underline{cww2104@columbia.edu}}}

\date {}
\maketitle

\begin{abstract}

We examine the cavity resonance tuning of high-$Q$ silicon photonic crystal heterostructures by localized laser-assisted thermal oxidation using a 532 nm continuous wave laser focused to a 2.5 $\mu$m radius spot-size.  The total shift is consistent with the parabolic rate law. A tuning range of up to 8.7 nm is achieved with $\sim$ 30 mW laser powers. Over this tuning range, the cavity $Q$ decreases from 3.2\e{5} to 1.2\e{5}. Numerical simulations model the temperature distributions in the silicon photonic crystal membrane and the cavity resonance shift from oxidation. \\

\end{abstract}

\section{Introduction}

Photonic crystal nanocavities are increasingly employed in photonic studies and applications because of their high quality factor ($Q$) to modal volume ($V_{m}$) ratios \cite{Noda05,Yang09,Gao10}. These nanocavities are also used in photonic devices of increasing complexity where high accuracy of the resonant wavelength is critical. However, due to fabrication imperfections, resonances will often deviate from their desired precise values. Several post-fabrication tuning techniques have been proposed and demonstrated to address this issue. These methods can be divided into 2 groups: global and local. Global tuning creates a uniform change over the entire chip, \cite{Yang07,Song09,Hennessy05} whereas local tuning only changes a small area such as a single nanocavity \cite{Lee09,Seo08,Faraon08,MWLee09,Hennessy06,Wong04,Pan08}. Global tuning is useful for correcting uniform errors but cannot address the random local errors that often occur during fabrication. Local tuning can be very important to applications such as the all-optical analog to electromagnetically induced transparency \cite{Yang09} and optical buffers, which require precisely coupled cavities.  Another example is in solid-state cavity quantum electrodynamics where local tuning can be used to spectrally match a cavity resonance to a single exciton transition.

\begin{figure}[htbp]
\centering\includegraphics[width=4.9in]{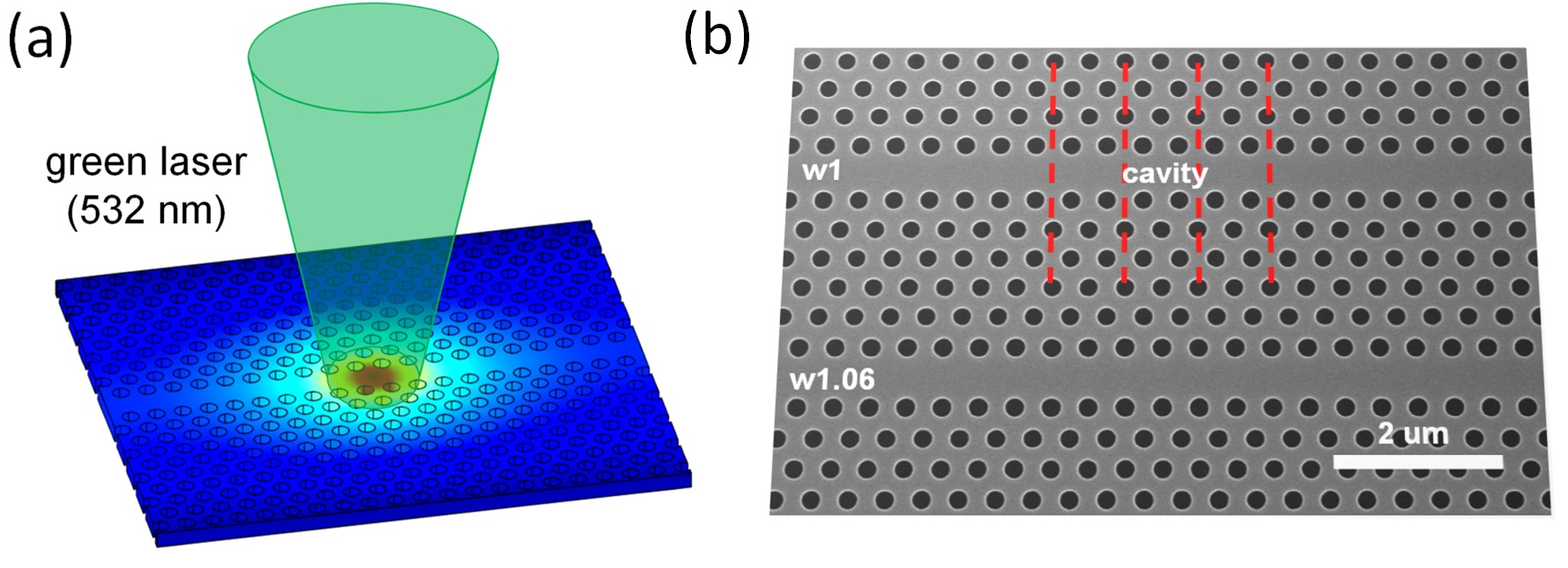}
\caption{\label{fig:fig1} (a) Illustration of laser-assisted local thermal oxidation on a silicon double-heterostructure cavity. (b) SEM image of a double-heterostructure cavity.
}
\end{figure}

In this work, we study the selective thermal oxidation of silicon photonic crystal membranes with a highly localized laser beam at ambient conditions in order to finely tune the resonances of high-$Q$ nanocavities (Fig. 1). Laser-assisted local oxidation is advantageous to previously demonstrated tuning techniques because it allows for automation of the tuning process by using a computer-controlled stage, shutter or optical modulator and in-situ monitoring system. Such a system would enable the post-fabrication fine-tuning of large numbers of nanocavities.

Laser-assisted local tuning has been demonstrated with the photodarkening of chalcogenide films on GaAs photonic crystals \cite{Faraon08} with a $Q$ of $\sim$ 8000. Laser-assisted local tuning has also been studied in the oxidation of GaAs photonic crystal L3 cavities \cite{Lee09} with a $Q$ of 1800. In this work, we study the precise local tuning in silicon, of photonic crystal double-heterostructure nanocavities with high-$Q$s of $\sim$ 300,000 or higher.

Continuous-wave lasers have been used in oxidation studies of silicon and silicon-on-insulator substrates \cite{Micheli87,Huber00}, including temperature independent contributions to silicon oxidation from photon flux \cite{Micheli87}. Using a diffraction limited beam, Deustchmann et al. \cite{Deutschmann99} was able to oxidize lines as narrow as 200 nm at a power of approximately 15 mW. Such spatial confinement is possible in thin single crystal silicon films because increased phonon scattering reduces heat flow in the lateral direction \cite{Ju99}. In addition to cavity tuning, local oxidation might be applicable to other post-fabrication tasks such as tuning the dispersion \cite{CChen10, MWLee07} and surface states \cite{Vlasov06,Chatterjee08}.

\section{Local oxidation cavity resonance tuning}

The photonic crystal double-heterostructure nanocavities \cite{Noda05} used in this work were fabricated by high quality photolithography and dry etching on silicon-on-insulator (250 nm thick) substrates, with 117 nm hole radii and 410 nm lattice parameter. In the cavity region, the lattice parameter increases to 415 and 420 nm. The waveguide-to-cavity separation is 6 layers of air-holes. Approximately 1.5 $\mu$m of oxide beneath the photonic crystal region was removed as described in Ref. 4.

A 532 nm diode-pumped solid-state laser, with a collimated beam and power controlled using a variable neutral density filter to $\sim$ 20 mW powers, was used. A 60$\times$ objective lens (NA of 0.65) focuses the laser onto the chip and is also used for imaging. The spot-size is measured using the knife-edge technique. The full-width half-maximum beam waist was 2.5 $\mu$m, corresponding to a maximum energy density at the chip surface of 1\e{8} W/cm$^{2}$. The oxidation is carried out in ambient conditions of $22 \,^{\circ}\mathrm{C}$ and 20\% relative humidity.

Cavity resonance transmission and radiation measurements were performed as described in Refs. 2 and 4. Oxidation results in a blueshift in the cavity resonance as shown in Fig. {\ref{fig:fig2} (a). This is the result of a larger decrease in refractive index from the silicon consumed outweighing a smaller increase in refractive index from the oxide generated. Following the initial anomalous oxidation, subsequent tuning follows a parabolic rate law where the cavity resonance blueshift is observed to be proportional to the square root of the oxidation time (Fig. {\ref{fig:fig2} (b) and (c)). The decreasing oxidation rate is attributed to the longer diffusion time of oxygen through the thicker oxide in order to reach the silicon-oxide interface where the oxide growth occurs \cite{DealGroveLaw}. Corresponding radiation measurements of the high-Q cavity are shown in Fig. \ref{fig:fig3}.

\begin{figure} [htbp]
\centering\includegraphics[width=4.5in]{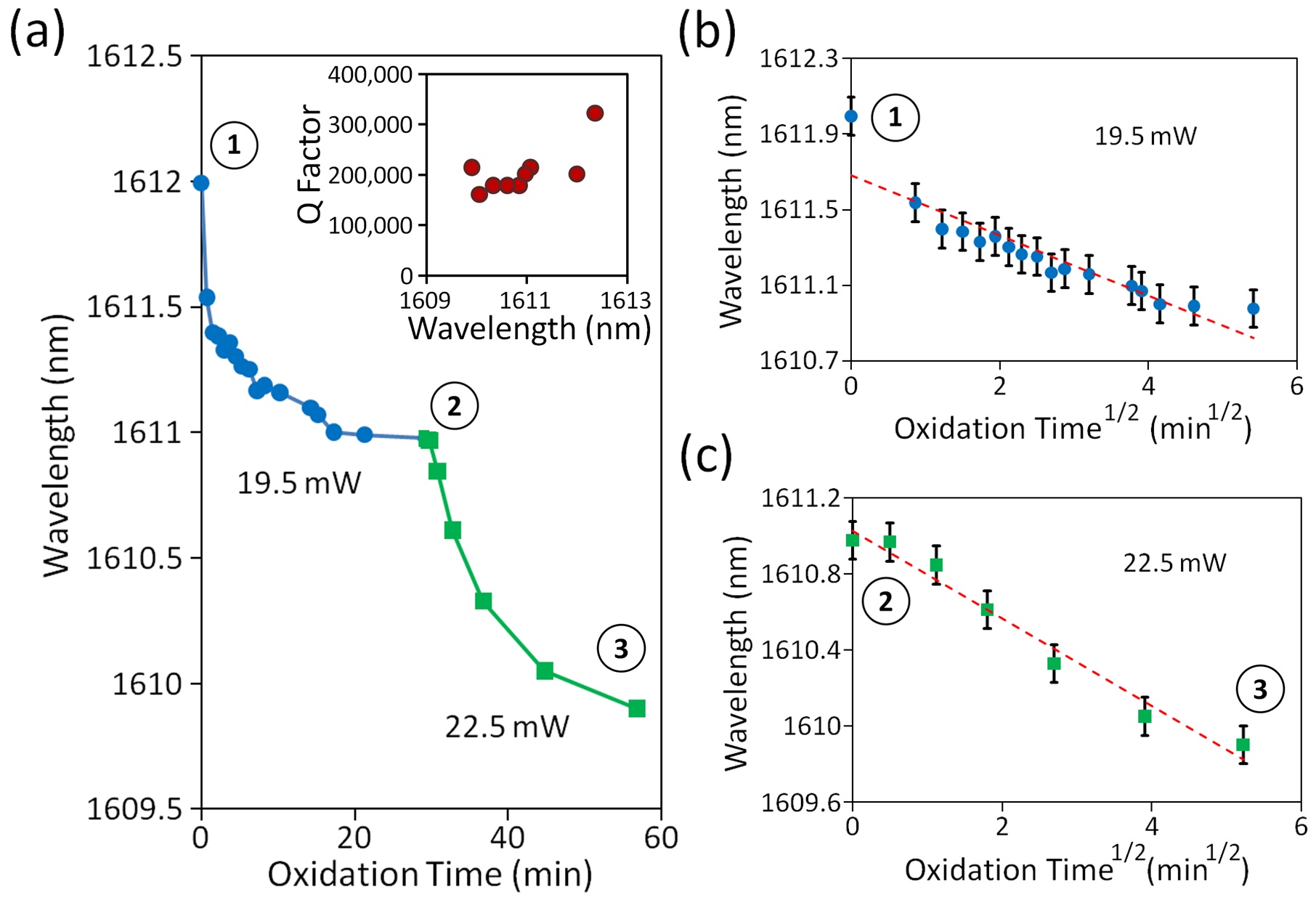}
\caption{\label{fig:fig2} (a) Experimental results showing the blueshift from tuning using an initial laser power of 19.5 mW (at the device surface). The same device is then further tuned at 22.5 mW. The inset shows measurements of the loaded quality factor as the cavity is tuned.  (b) Fitting of the resonant wavelength shift to the square root of oxidation time for incident power of 19.5 mW and (c) for 22.5 mW.}
\end{figure}

\begin{figure}[htbp]
\centering\includegraphics[width=3.3in]{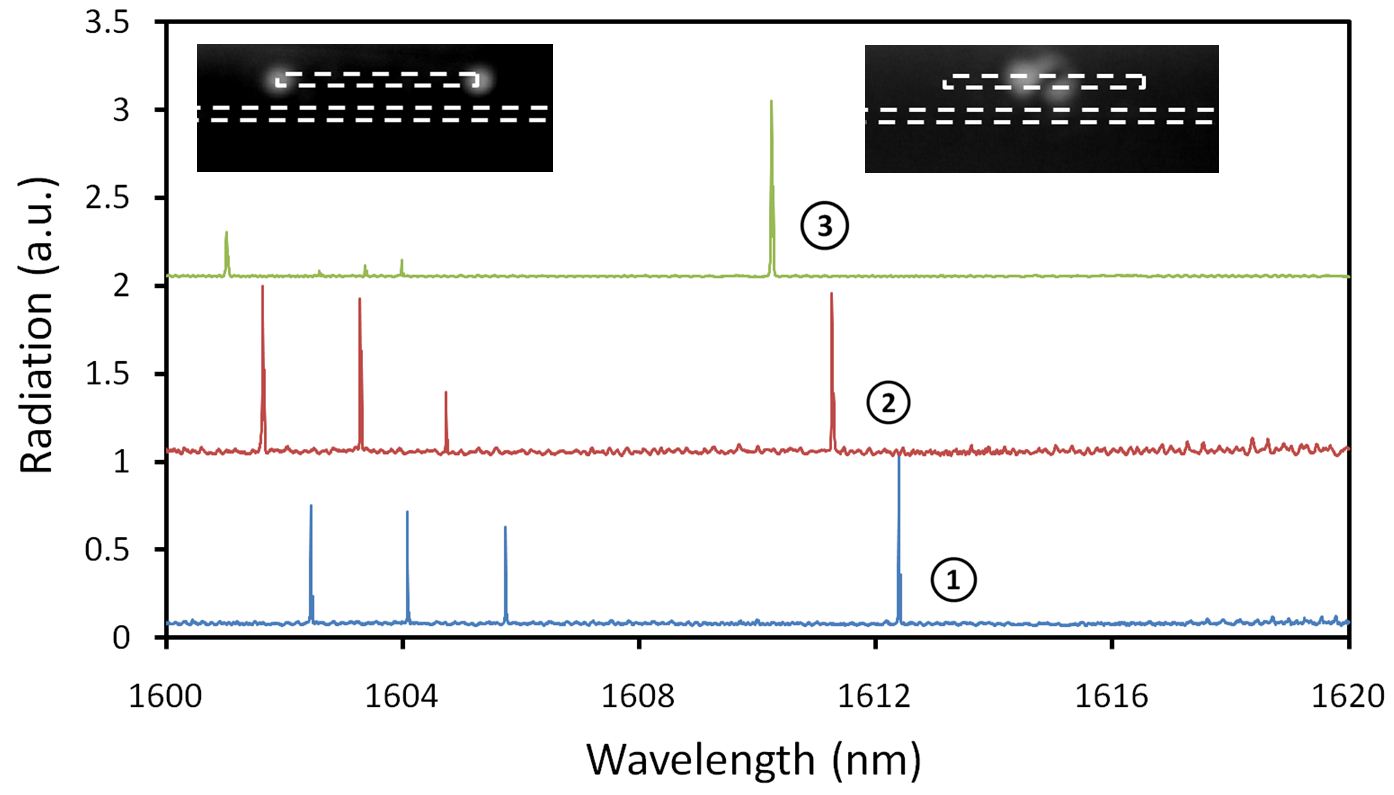}
\caption{\label{fig:fig3} Radiation measurements at three different tuning increments corresponding to the numbered positions in Fig. 2. Left inset: near-infrared radiation pattern for the Fabry-Perot modes on the left. The dotted lines indicate the position of cavity waveguide (upper) and input/output waveguide (lower). Right inset: near-infrared radiation pattern for the cavity mode on the right-side of the spectra.}
\end{figure}

Oxidation is expected to occur on both the top and bottom surfaces of the silicon membrane as well as the surface of the air holes because both the 1/e absorption depth of the green laser (1 $\mu$m) and the depth of focus of the beam after the objective lens (400 nm for 60$\times$ objective) are larger than the 250 nm thickness of the silicon membrane. The oxidation profile remains highly localized with AFM studies showing lateral dimensions 40\% smaller than the spot diameter \cite{Huber00}. This can be attributed to a number of factors. The thermal conductivity of silicon is dominated by phonon transport with a smaller contribution from free charge carriers. While the thermal conductivity of silicon is relatively high at room temperature, it decreases significantly at higher temperatures with increased phonon scattering \cite{Shanks63,Asheghi98}. In addition, as the substrate thickness decreases from bulk dimensions, phonon-boundary scattering increases and the presence of air holes will further decrease the phonon mean free path \cite{Ju99}.

\section{Transient effects from oxide surface chemistry}

During the oxidation process, as the cavity is heated to high temperatures by optical absorption of the focused green laser beam, there will be a large redshift in the cavity resonance. The redshift is attributed to the thermal-optic effect in silicon causing a temperature-dependent refractive index change. When the green laser is turned off, the cavity temperature quickly returns to room temperature and the large redshift disappears. But within 1-2 minutes after oxidation (i.e. after the green laser is turned off), a smaller magnitude redshifting of the cavity resonance is observed. The rate of this redshifting is rapid at first and decreases over time, taking many hours to reach a stable value. This transitory effect is caused by water molecules on the surface of the cavity.

The surface chemistry effects occur both during and after laser irradiation, and are illustrated in Fig. \ref{fig:fig4}. Upon heating the cavity there is a temporary blueshift from oxide surface and oxide bulk dehydration. After the cavity cools back down, there is a slow redshift caused by gradual rehydration of water molecules onto the cavity. The total shift ranges from $\sim$ 100 pm to hundreds of picometers and appears to be dependent on the oxide thickness. Laser exposure at low powers (less than 1 mW for 10 minutes) shows a completely reversible blueshift, indicating no real oxidation has occurred (i.e. there is only a temporary blueshift from cavity dehydration followed by a rehydration redshift returning the cavity resonance to its original wavelength).

\begin{figure}[htbp]
\centering\includegraphics[width=3.45in]{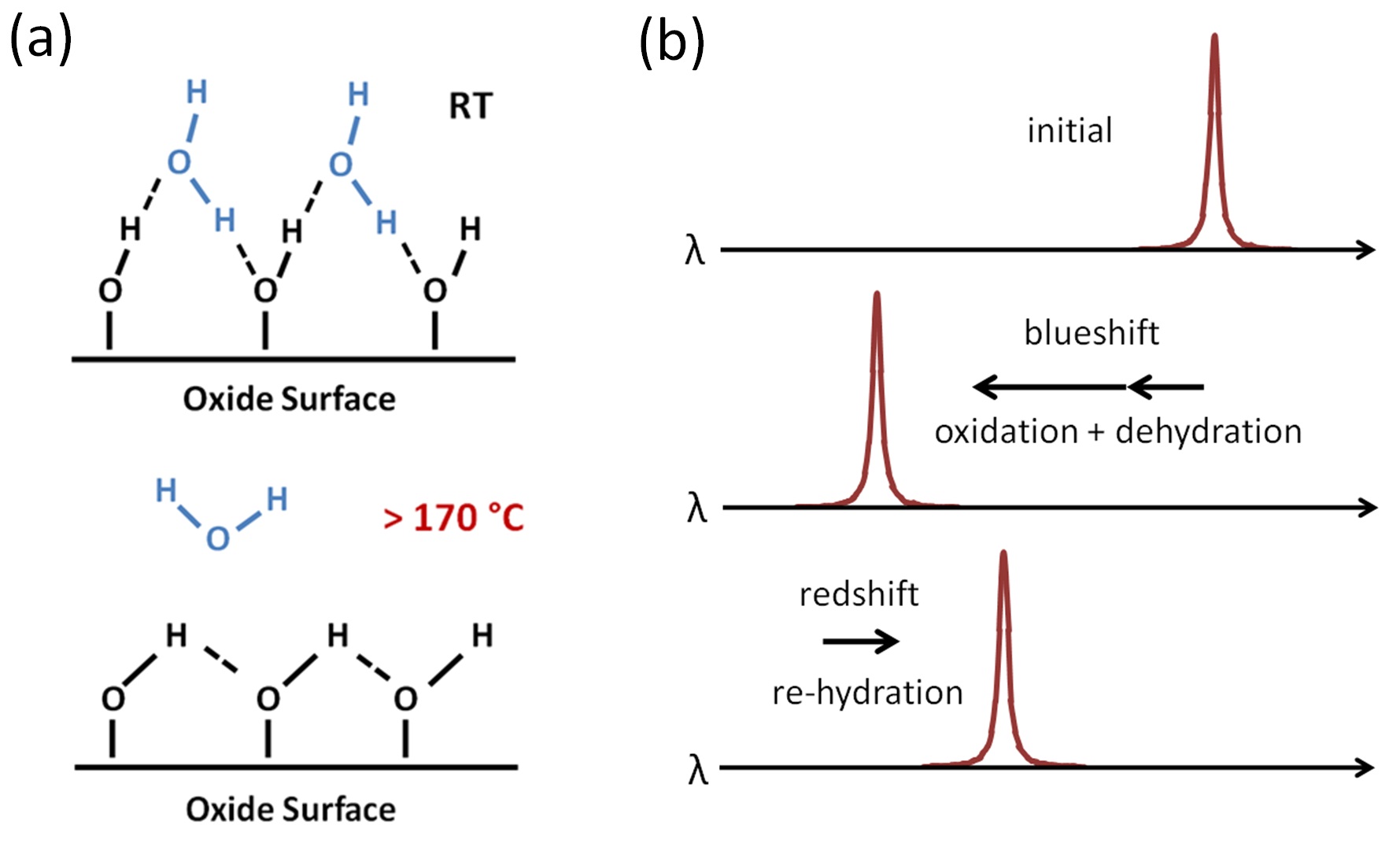}
\caption{\label{fig:fig4} Transitory surface chemistry effects. (a) Water molecules absorb onto the oxide surface at room temperature and desorb from the oxide surface at elevated temperatures. (b) During laser irradiation, the cavity experiences a resonance blueshift from oxide dehydration in addition to silicon oxidation. After the cavity cools, water will slowly rehydrate the oxide surface resulting in a gradual redshift.}
\end{figure}

Silicon dioxide terminated by hydroxyl groups (SiOH) is hydrophillic and will readily adsorb water molecules, as schematically shown in Fig. ~\ref{fig:fig4} (a). At temperatures above $\sim$ 170$\,^{\circ}\mathrm{C}$ the hydrogen bonded water molecules will desorb. The process is reversible but at higher temperatures ($400 \,^{\circ}\mathrm{C}$) the hydrogen in the hydroxyl groups can sometimes be removed resulting in a hydrophobic siloxane surface. The long times required to complete the redshift are indicative of a slower diffusion-limited process \cite{SilicaBook, LeGrange93}. While water absorption is known to be much less significant for thermally grown oxides than deposited oxides \cite{DeRooij77}, studies of laser grown oxides indicate their composition are less dense than traditional thermally oxidized films due to presence of suboxides, especially for thinner films \cite{Aygun05}. In order to obtain reliable results, our measurements were taken either immediately after oxidation (less than 1 min) when the cavity region was still dehydrated or immediately after the cavity was re-heated at sub-oxidation threshold powers.

\section{Thermal oxidation}

Prior to oxidation, the test chip had a native oxide (approximately 12 \AA\ in thickness) as it was exposed to ambient conditions. In addition to the native oxide, we observed that initial oxidation can occur at very low laser powers. Because of this, an accurate initial oxidation threshold power was difficult to determine. This initial growth was anomalous in respect to the Deal-Grove and Massoud models for oxide growth \cite{DealGroveLaw, Massoud85}. It has been shown by x-ray photoelectron spectroscopy studies that the initial $\sim$ 22 \AA\ of oxide growth occurs at a very high rate \cite{Enta08}. At this stage, even at very low powers oxidation might occur.

Following the initial anomalous oxidation, subsequent tuning follows a parabolic rate law. The oxidation resonance tuning is permanent and stable. Repeated measurements over the span of many days shows no change within the measurement error. The uncertainty in the cavity resonance measurements is approximately $\pm$ 100 pm and primarily attributed to random thermal fluctuations in the ambient environment. This uncertainty is the limiting factor in determining the minimum tuning increment. A number of other factors will affect the minimum achievable tuning increment including: power stability, exposure time accuracy, beam targeting accuracy and beam shape uniformity. Another group has demonstrated a resonance tuning increment of 100 pm using atomic force microscope induced nano-oxidation of GaAs \cite{Hennessy06}.

\begin{figure}[htbp]
\centering\includegraphics[width=5.6in]{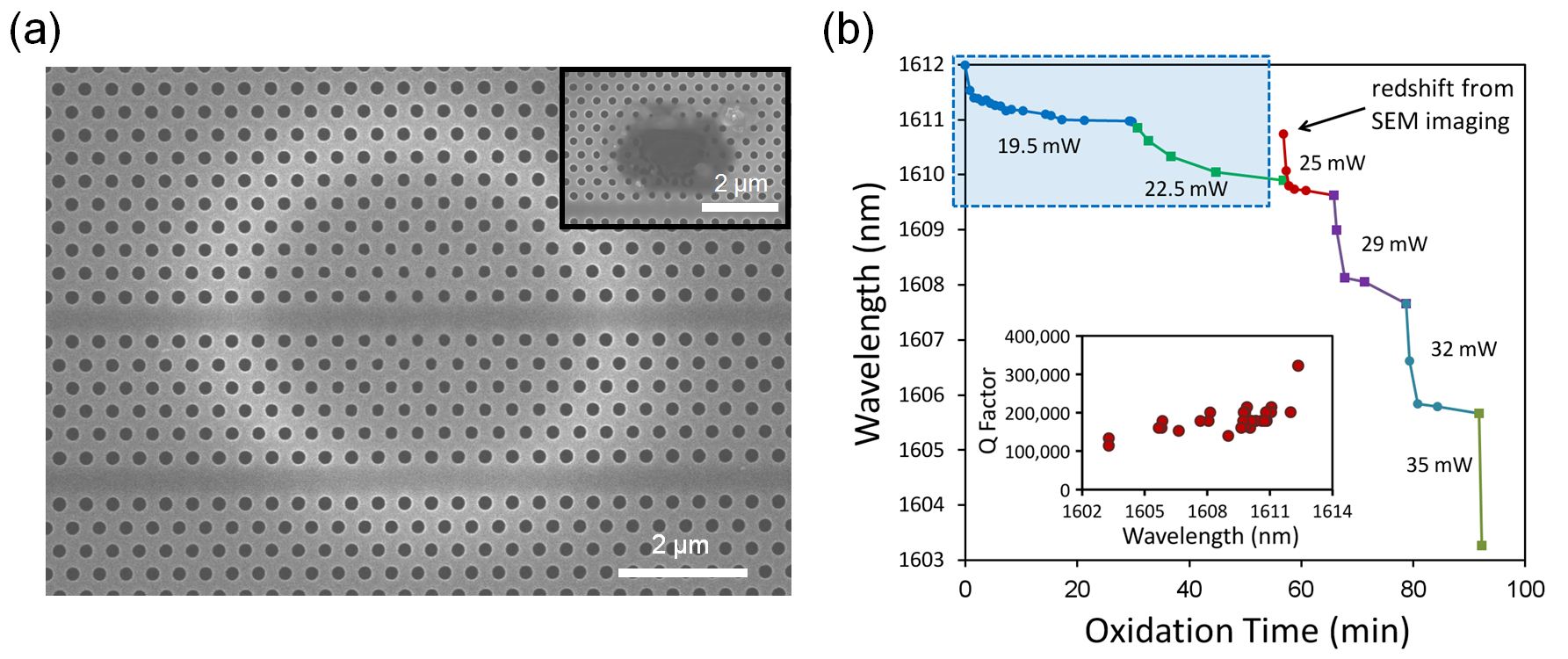}
\caption{\label{fig:fig5} (a) SEM image of cavity after local oxidation tuning at laser power of 35 mW. The white ring indicates oxide charging effects during SEM imaging.  The center region is darker because of slight melting. Inset: Hole melted through the silicon membrane after irradiation at $\sim$ 35 mW for several minutes (different device from the main figure). (b) Local oxidation tuning over a larger wavelength range. The blue region (upper-left) corresponds to the data shown in Fig. 2 (a). The upper-right arrow corresponds to the SEM image in Fig. 1 (b). The inset shows measurements of the loaded quality factor.}
\end{figure}

The inset to Fig. \ref{fig:fig2} (a) shows quality factor measurements during the oxidation process. During the initial oxidation there is a sizeable drop in the Q. Afterwards the Q appears to remain stable upon further oxidation. The initial drop in Q might be attributed to increased optical scattering from the interfacial layer between the silicon and oxide. Surface states might also result in increased absorption at the interface. In this work, the local oxide was left on the chip after the tuning. Removing the oxide with an HF dip \cite{Lee09} has been shown to improve the Q and will also further blueshift the resonance.

The decrease in overall air-hole size can be used to estimate the oxide thickness. The SEM image in Fig. \ref{fig:fig1} (b) shows the double-heterostructure cavity after local oxidation resulting in a 2.1 nm cavity resonance blueshift. Image analysis of the experimental device indicates a hole size change smaller than the error of the SEM measurements ($\pm$ 5\% of the hole diameter). In addition, a redshift of 0.8 nm was observed in the cavity resonance after SEM imaging the cavity region (Fig. \ref{fig:fig5} (b)). This is likely caused by surface contamination of the sample in the SEM chamber during imaging. It is well known that organic molecules from vacuum pump oil can be deposited onto the substrate surface by the focused electron beam during SEM imaging \cite{Seo08}. For this reason, SEM imaging was generally avoided while the sample was being tuned. Fig. \ref{fig:fig5} (a) shows an SEM image of the sample after cavity resonance tuning over the full range shown in Fig. \ref{fig:fig5} (b). The total resonance blueshift is 8.7 nm. Note that if the SEM-induced contamination redshift did not occur, then the total resonance blueshift would be 9.5 nm. Using hole-size analysis it is estimated that 1 nm of resonance blueshift corresponds to 1.9 nm of oxide growth. A comparison with calculated resonance shifts is provided in the next section.

The measured loaded $Q$ of the cavity over the entire tuning range decreases from 3.2\e{5} to 1.2\e{5} as shown in the inset of Fig. \ref{fig:fig5} (b). It has been shown by 3D FDTD analysis that, for the double-heterostructure cavity design used, an increase in hole radii of $\sim$ 12\% corresponds to a decrease in $Q$ of $\sim$ 80\% \cite{Noda05}. In our experimental case, our hole radii increases by $\sim$ 8\% and Q decreases by $\sim$ 63\% (only considering the silicon; note that the overall hole radii actually decreases because of the oxide growth). Hence, the sensitivity of $Q$ to the hole size plays a significant role in the observed drop in $Q$ during tuning.

\section{Numerical analysis}

Three-dimensional finite-element simulations (COMSOL Multiphysics) are used to estimate the temperature increases caused by optical absorption at different laser powers. The model uses the static heat equation:
\begin{equation}\label{first}
\nabla [\kappa (T) \nabla T] = -[1-R(T)]I_0\alpha(T)\exp \left( -\frac{(x^2+y^2)}{2\sigma^2} \right) \exp \left( -\int_0^z \alpha(T(x,y,z'))\mathrm{d}z'\right)
\end{equation}
where $\kappa(T)$ is the temperature-dependent thermal conductivity, $R(T)$ is the reflection coefficient and $\alpha(T)$  is the temperature-dependent absorption coefficient. The incident CW laser beam has a Gaussian spatial distribution with a 1/e-squared beam spot-size $\sigma$ and a center intensity $I_0$. The reflection coefficient takes the form   $R(T) = R_0 + c(T-T_0)$ \cite{Liarokapis85,Micheli87}. The temperature dependent absorption is experimentally fitted as $\alpha(T) = \alpha _0 \exp(T/T_0)$  \cite{Jellison82}. Both the absorption and thermal conductivity of silicon are temperature dependent. The absorption of silicon increases significantly at higher temperatures. The thermal conductivity for bulk silicon at room-temperature is 148 WK$^{-1}$m$^{-1}$ but as the temperature increases, the thermal conductivity will decrease. For a thin porous photonic crystal slab, the thermal conductivity can be reduced even further. In-plane thermal conductivity is reduced to approximately 90 WK$^{-1}$m$^{-1}$ at room temperature for silicon-on-insulator (SOI) devices with a silicon thickness of 260 nm \cite{Aubain10}, mainly due to increased phonon boundary scattering \cite{Asheghi98,Song04,Hopkins09}. Periodic structures on single-crystalline silicon membranes can have thermal conductivity values as small as $\sim$ 6.8 WK$^{-1}$m$^{-1}$ due to coherent phononic effects \cite{Hopkins11}.

In order to determine the thermal conductivity for our device, we used our finite-element model along with experimental measurements. Silicon melting is observed to occur at incident laser powers of $\sim$ 35 mW as shown in Fig. \ref{fig:fig5} (a). This was used as a reference point for our model and the resulting room-temperature thermal conductivity was estimated to be 67 WK$^{-1}$m$^{-1}$. The simulated temperature profile around the double-heterostructure cavity is shown in Fig. \ref{fig:fig6} (a). Cross-sections of the temperature profile corresponding to incident power of 30 mW and 35 mW are shown in Fig. \ref{fig:fig6} (b) along with the laser intensity profiles at those powers. The temperature profiles show how the temperature gradient increases due to higher absorbed energies and lower thermal conductivity at higher membrane temperatures. The local maximum temperature for different incident laser powers is shown in Fig. \ref{fig:fig6} (c). The features observed in Fig. \ref{fig:fig5} (a) are consistent with results of the numerical model Fig. \ref{fig:fig6} (a).

\begin{figure}[htbp]
\centering\includegraphics[width=5.4in]{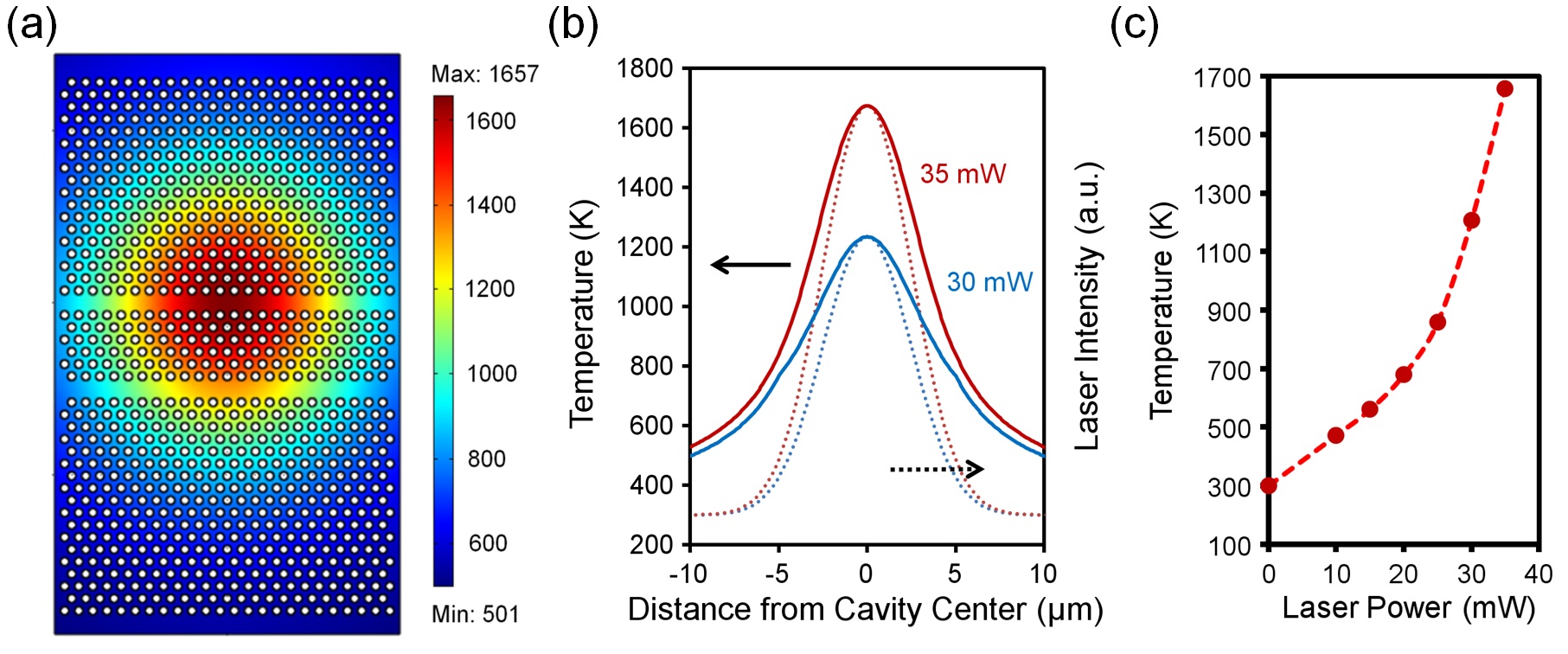}
\caption{\label{fig:fig6} (a) Finite-element simulation (COMSOL Multiphysics) of temperature distribution across silicon double-heterostructure cavity during 532 nm laser irradiation at 35 mW. (b) Solid lines represent the temperature distribution as a function of distance from the center of the laser beam. Dotted lines represent the intensity profile of the laser beam. (c) Simulation results of local maximum temperature versus laser power. Temperatures range from room temperature to the melting point of silicon.
}
\end{figure}

\begin{figure}[htbp]
\centering\includegraphics[width=5.2in]{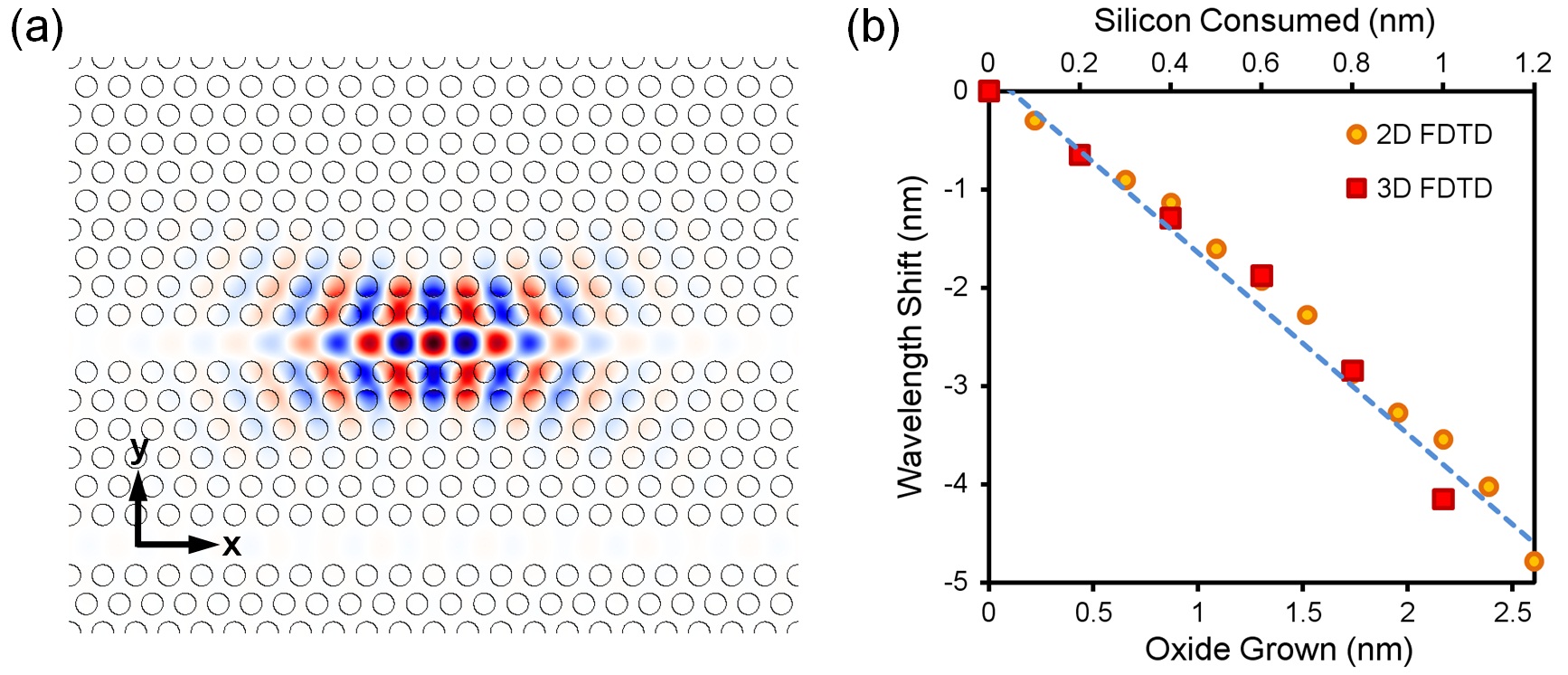}
\caption{\label{fig:fig7} (a) Calculated electric field $E_{y}$ profile of high-$Q$ mode supported by double-heterostructure cavity. (b) Calculated wavelength shift of the resonant mode due to local oxidation of the silicon photonic crystal membrane.
}
\end{figure}

The finite-difference-time-domain (FDTD) method \cite{Oskooi10} with sub-pixel averaging \cite{Hagino09} is used to calculate the effects of silicon oxidation on the cavity resonance (Fig. 7). The native oxide is assumed to be 1.2 nm. We also assume that if 1 nm of silicon is oxidized (consumed), the resulting oxide grown is 2.17 nm (0.46 ratio). As the silicon is consumed, the total thickness of the slab (silicon + oxide) will increase, while the air-hole radii will decrease. Simulation results in Fig. \ref{fig:fig7} (b) show that every 0.6 nm of oxide grown results in 1 nm of resonance blueshift. By comparison, the previously discussed SEM hole analysis estimates that 1.9 nm of oxide growth results in 1 nm of resonance blueshift. There are several possible explanations for this discrepancy. It is possible that the amount of oxidation on the bottom-side of the silicon membrane is less than the top-side if the beam focus is less than optimal. In addition, the laser grown oxide might be less dense than furnace grown thermal oxide \cite{Aygun05}. Finally, the slight melting of the silicon membrane at 35 mW laser irradiation might have resulted in a decrease in the air-hole size without a proportional increase in oxide growth.

\section{Conclusion}

We have demonstrated the tuning of high-$Q$ double-heterostructured silicon photonic crystal nanocavities using laser-assisted local thermal oxidation. Cavity $Q$ decreases from 3.2\e{5} to 1.2\e{5} over the range of oxidation times and laser powers examined. The effects of water absorption and thin oxide growth were also observed. Numerical simulations were used to model the temperature distribution in the silicon photonic crystal membrane and resonance shift of the optical mode due to oxidation.
\\

\section{Acknowledgements}

The authors acknowledge funding support from DARPA DSO and NSF ECCS. Work was also performed at the Columbia University Cleanroom, which is supported by the NSEC Program of the National Science Foundation under Award Number CHE-0641523 and the New York State Foundation for Science, Technology, and Innovation (NYSTAR).\\

\end{document}